\documentclass[preprints,accept,submit,moreauthors,pdftex]{mdpi} 

\firstpage{1} 
\pubvolume{1}
\issuenum{1}
\articlenumber{0}
\pubyear{2021}
\copyrightyear{2020}
\datereceived{} 
\dateaccepted{} 
\datepublished{} 
\hreflink{https://doi.org/} % If needed use \linebreak
\Title{Improving the Authentication with Built-in Camera Protocol Using Built-in Motion Sensors: A Deep Learning Solution}

\TitleCitation{Improving the Authentication with Built-in Camera Protocol Using Built-in Motion Sensors: A Deep Learning Solution}

\Author{Cezara Benegui $^{1}$\orcidA{}, and Radu Tudor Ionescu\orcidB{} $^{2}$}

\AuthorNames{Cezara Benegui, Radu Tudor Ionescu}

\AuthorCitation{Benegui, C.; Ionescu, R.T.;}
\address{%
$^{1}$ \quad Faculty of Mathematics and Computer Science, University of Bucharest, Romania; cezara.benegui@fmi.unibuc.ro\\
$^{2}$ \quad Faculty of Mathematics and Computer Science / Romanian Young Academy, University of Bucharest, Romania; raducu.ionescu@gmail.com}

\firstnote{This article is an extended version of our paper published at the 18th International Conference on Applied Cryptography and Network Security (ACNS 2020).} 
\abstract{In this paper, we propose an enhanced version of the Authentication with Built-in Camera (ABC) protocol by employing a deep learning solution based on built-in motion sensors. The standard ABC protocol identifies mobile devices based on the photo-response non-uniformity (PRNU) of the camera sensor, while also considering QR-code-based meta-information. During registration, users are required to capture photos using their smartphone camera. The photos are sent to a server that computes the camera fingerprint, storing it as an authentication trait. During authentication, the user is required to take two photos that contain two QR codes presented on a screen. The presented QR code images also contain a unique probe signal, similar to a camera fingerprint, generated by the protocol. During verification, the server computes the fingerprint of the received photos and authenticates the user if $(i)$ the probe signal is present, $(ii)$ the metadata embedded in the QR codes is correct and $(iii)$ the camera fingerprint is identified correctly. However, the protocol is vulnerable to forgery attacks when the attacker can compute the camera fingerprint from external photos, as shown in our preliminary work. Hence, attackers can easily remove their PRNU from the authentication photos without completely altering the probe signal, resulting in attacks that bypass the defense systems of the ABC protocol. In this context, we propose an enhancement for the ABC protocol, using motion sensor data as an additional and passive authentication layer. Smartphones can be identified through their motion sensor data, which, unlike photos, is never posted by users on social media platforms, thus being more secure than using photographs alone. To this end, we transform motion signals into embedding vectors produced by deep neural networks, applying Support Vector Machines for the smartphone identification task. Our change to the ABC protocol results in a multi-modal protocol that lowers the false acceptance rate for the attack proposed in our previous work to a percentage as low as $0.07\%$. In this paper, we present the attack that makes ABC vulnerable, as well as our multi-modal ABC protocol along with relevant experiments and results.}


\begin{document}
%%%%%%%%%%%%%%%%%%%%%%%%%%%%%%%%%%%%%%%%%%

\section{Introduction}
\label{sec:introduction}

Rapid advancement of mobile device technology, such as development of high-resolution cameras, contributes to a large volume of data shared across the World Wide Web through social media platforms and other online environments. Moreover, smartphones are now the medium of choice for accessing applications that require strong security, such as banking applications~\cite{Mobile-FRS-2016}, thus making them ideal targets for attackers. While any device carries a minimum of security mechanisms, the information generated by mobile devices can be used in identity forgery attacks~\cite{arthur2013iphone} or in side-channel attacks such as smudge~\cite{Aviv-WOOT-2010} and reflection~\cite{Xu-CCS-2013,Zhang-SPSM-2012}. To defend against attacks, Zhongjie et al.~\cite{Zhongjie-NDSS-2018} proposed the Authentication with Built-in Camera (ABC) protocol, based on a special characteristic of the camera sensor, namely the photo-response non-uniformity (PRNU)~\cite{Lukas-TIFS-2006}. The PRNU fingerprint is represented by a noise pattern unique to each camera sensor, which can be determined as detailed in previous  research~\cite{Lukas-TIFS-2006,Altinisik-EUSIPCO-2018, Akshatha-DI-2016,Li-TIFS-2010,Cooper-FSI-2013,Kang-TIFS-2012}, even from a single photo. The ABC protocol introduced by Zhongjie et al.~\cite{Zhongjie-NDSS-2018} uses the camera fingerprint as the main authentication factor and is composed of two phases, particularly a registration phase in which the PRNU fingerprint of the device is computed and stored, and secondly, an authentication phase. During authentication, a registered device takes photos of two QR codes presented on a screen, and sends them to a server for identification. The server performs a set of tests consisting of QR code metadata validation, camera fingerprint identification and forgery detection.  

Photos uploaded on social media platforms offer attackers the possibility to compute PRNU fingerprints of potential victims with the aim of conducting impersonation attacks. To prevent this possibility, in addition to the QR codes, the photos received by the ABC server contain a probe signal, represented by a noise pattern similar to a camera fingerprint. The ABC forgery detection system can determine if an attacker tries to impersonate a legitimate user by testing if the probe signal is missing from the provided QR code images. Zhongjie et al.~\cite{Zhongjie-NDSS-2018} concluded that, in the process of replacing the attacker's fingerprint with the victim's fingerprint, the probe signal is also removed. The ABC protocol supposes that attackers compute fingerprints during the authentication phase, from the photos containing the QR codes. However, as we further detail in this work, the protocol is susceptible to an attack that can easily bypass the ABC forgery detection system.

We present an attack strategy, initially introduced in our preliminary work~\cite{benegui2020breach}, in which the attacker computes their fingerprint from any other set of photos taken with the device involved in the attack. Hence, when the attacker's fingerprint is removed from the QR code photos and replaced with the victim's fingerprint, the probe signal is no longer removed. In this scenario, the forgery detection system will not identify the attacker, rendering the protocol vulnerable to attacks. %The probe signal is significantly similar to a camera fingerprint, thus, during the replacement procedure carried by the adversarial, it is altered to some extent. 
%Therefore, 
The proposed attack has a success rate of around $50\%$. Nevertheless, in this work, we propose a novel method of enhancing the ABC protocol that can prevent the attack strategy described in~\cite{benegui2020breach}, increasing the overall security level. Our proposal is based on considering motion signals, namely those provided by the gyroscope and the accelerometer sensors, which are typically built-in components of modern smartphones. This results in a passive two-factor (multi-modal) authentication protocol that reduces the attack success rate to $0.07\%$. 

%Discrete signals generated by the accelerometer and gyroscope are recorded during the tap gesture on the smartphone screen, starting with 0.5 seconds prior to the tap event and continuing for another 1 second. The recording is performed when the user captures photos of the two QR codes during the authentication phase in the ABC protocol. Values on each axis (x, y, z) reported at a sample rate of 100 Hz, produce 150 values per axis, in a time frame of 1.5 seconds. Hence, for each tap gesture we obtain 6 discrete signals. Based on Benegui et al.~\cite{Benegui-Access-2020} we transform the signals into gray-scale image representation, following a modified de Brujin~\cite{Ralston-MM-1982} sequence. Furthermore, the obtained images are passed to deep convolutional neural networks (CNNs)~\cite{LeCun-IEEE-1998,Krizhevsky-NIPS-2012} trained for few-shot user identification, used as feature extractors. Likewise, the motion sensor data is provided as input to Conv-LSTM, trained for the same purpose and used as a feature extractor. The resulted embeddings are forwarded as input to Support Vector Machines (SVMs)~\cite{Cortes-ML-1995} with the aim to identify the user. Hence, our extended work adds another two factor authentication system to the original protocol.

To learn motion signal patterns specific to the built-in sensors, we propose a deep learning approach that combines features from deep convolutional neural networks (CNNs)~\cite{LeCun-IEEE-1998,Krizhevsky-NIPS-2012} and convolutional Long Short-Term Memory networks (ConvLSTMs)~\cite{Xingjian-NIPS-2015} using an ensemble model based on Support Vector Machines (SVMs)~\cite{Cortes-ML-1995}. To demonstrate the effectiveness of our improved ABC protocol, we perform multi-modal experiments using 630 images and corresponding motion signals collected from six different mobile devices. %Zhongjie et al.~\cite{Zhongjie-NDSS-2018} use one photo during registration phase, while we use 5 photos, such that the PRNU estimation is more robust. We note that using multiple photos, namely 5, both for the attacker and the victim to compute the fingerprint, the attack is more likely to succeed. However, our additional authentication layer based on motion sensor data, considerably lowers the success rate of the attack, even in this scenario. 
From the 105 examples per device, we use 5 examples for registration and the rest of 100 examples for authentication experiments. Each device becomes a victim of impersonation attacks throughout our experiments, employing a total of 500 attack sessions from the other five devices. Our experiments on the enhanced multi-modal ABC protocol, consisting in a total of 3000 attacks, indicate a false acceptance rate (successful attack rate) of $0.07\%$. Compared to the original ABC protocol~\cite{Zhongjie-NDSS-2018}, which has a false acceptance rate of $54.1\%$ in the same scenario, our protocol enhancement results in an improvement of $45.83\%$. % We note that, this improvement is obtained using the combined features extracted from both LSTM and CNN models, applied on the motion sensor data as a two factor authentication system on top of the original ABC protocol.  
We thus conclude that the multi-modal ABC protocol, enhanced with the authentication based on motion sensors, is more secure, becoming a promising candidate for smartphone device authentication. We emphasize that our additional authentication mechanism based on motion sensors is passive (or implicit), resulting in a seamless experience for the user.

The rest of this paper is organized as follows. We present related work in Section~\ref{sec_RelatedWork}. We describe the ABC protocol, its vulnerability and our enhanced multi-modal ABC protocol in Section~\ref{sec_Method}. Our experiments and results are presented in Section~\ref{sec_Experiments}. We draw our conclusions in Section~\ref{sec_Conclusion}.

\section{Related Work}
\label{sec_RelatedWork}

\subsection{Attacks on Authentication Protocols}

A recent work~\cite{Aghili-IACR-2018} proposed attack schemes for machine-to-machine (M2M) authentication protocols~\cite{Esfahani-ITJ-2019}. The authors showed that the M2M authentication protocol~\cite{Esfahani-ITJ-2019} is vulnerable to Denial-of-Service (DoS) and router impersonation attacks.
In a different work, Aghili et al.~\cite{Aghili-Sensors-2018} showed that the untraceable and anonymous three-factor authentication scheme~\cite{Amin-JNCA-2018} for Heterogeneous Wireless Sensor Networks is vulnerable to user impersonation, de-synchronization and traceability attacks. Aghili et al.~\cite{Aghili-Sensors-2018} also proposed an improved protocol that is resilient to these kinds of attacks.

In literature, the use of camera PRNU fingerprint in authentication systems or protocols is studied in a few works~\cite{Zhongjie-NDSS-2018,DBLP:conf/mediaforensics/AmeriniBBCCT16,DBLP:journals/tifs/ValsesiaCBM17}. Zhongjie et al.~\cite{Zhongjie-NDSS-2018} proposed a protocol based solely on PRNU fingerprint identification, while others~\cite{DBLP:conf/mediaforensics/AmeriniBBCCT16,DBLP:journals/tifs/ValsesiaCBM17} integrated the PRNU fingerprint in multi-factor authentication systems.
%, such as the ABC protocol described by Zhongjie et al.~\cite{Zhongjie-NDSS-2018}, authentication system based on multiple device sensor fingerprints (including PRNU)~\cite{DBLP:conf/mediaforensics/AmeriniBBCCT16} and Valsesia et al.~\cite{DBLP:journals/tifs/ValsesiaCBM17} multifactor authentication system that embeds the PRNU.
However, PRNU-based authentication protocols, such as the ABC protocol studied by Zhongjie et al.~\cite{Zhongjie-NDSS-2018}, have not been thoroughly studied in the presence of attacks. Zhongjie et al.~\cite{Zhongjie-NDSS-2018} introduced different attack prevention mechanisms, such as an anti-forgery detection system, but the protocol can still be breached through the attack scheme proposed in~\cite{benegui2020breach}. Moreover, it is known that camera fingerprints are vulnerable to attacks, e.g.~forgery attacks~\cite{gloe2007can,goljan2011defending}, thus rendering fingerprints unsafe as a single authentication factor. Hence, in this paper, we propose a multi-modal seamless extension of the ABC protocol. Instead of relying only on the camera fingerprint, we propose to add the joint fingerprint of two additional sensors: the gyroscope and the accelerometer.

\noindent
\textbf{Relation to preliminary ACNS 2020 version~\cite{benegui2020breach}.}
Since our current work is an extension of our previous paper~\cite{benegui2020breach}, we explain the differences in detail. In our previous work~\cite{benegui2020breach}, we discovered a breach in the design and implementation of the ABC protocol~\cite{Zhongjie-NDSS-2018}. We provided evidence of an attack scheme that shows how forgery attacks (adversary fingerprint removal) attacks affect the protocol. Different from our preliminary study~\cite{benegui2020breach}, we introduce an implicit authentication system based on motion sensor data in the ABC protocol, with the aim of improving the overall accuracy and stopping potential forgery attacks. Although in our previous work we tried to model motion sensor data using statistical features~\cite{benegui2020breach}, the accuracy of the resulting model was far below the requirements of an authentication protocol. Hence, in this work, we propose to employ an ensemble based on neural embeddings derived from two types of deep neural networks, CNNs and ConvLSTMs, concatenated and forwarded as input to an SVM meta-model. To our knowledge, we are the first to propose an ensemble of CNNs and ConvLSTMs for smartphone identification based on motion sensor data. Our improvements lead to a more robust ABC protocol, attaining superior performance in adversarial detection over the preliminary works~\cite{benegui2020breach,Zhongjie-NDSS-2018}.

\subsection{Authentication based on Motion Sensors}

Several papers studied user identification on mobile devices based on motion sensor data \cite{Buriro-SPW-2016,Buriro-ISBA-2017,Ku-Access-2019,Li-BIBM-2018,Neverova-Access-2016,Shen-Sensors-2016,Sitova-TIFS-2016,Sun-ECML-2017,Vildjiounaite-ICPC-2006,Wang-Access-2019}. Within the wide range of explored approaches, there are studies that perform user recognition based on voice and accelerometer signals \cite{Vildjiounaite-ICPC-2006}, as well as studies that perform human movement tracking based on motion sensors \cite{Sitova-TIFS-2016}. Regarding the considered approach, it is clear that the newest and best-performing methods belong to the category of deep learning approaches \cite{Benegui-Access-2020,Neverova-Access-2016}. Until now, researchers studied recurrent neural networks~\cite{Neverova-Access-2016} and convolutional neural networks~\cite{Benegui-Access-2020}. To our knowledge, none of the previous works investigated ensemble methods that combine recurrent and convolutional neural networks. We introduce an ensemble that uses an SVM as meta-learner. Different from previous works, we interpret the weights of the meta-learner as a joint fingerprint of the gyroscope and the accelerometer sensors.

%\vspace{-0.2cm}
\section{Method}
\label{sec_Method}
%\vspace{-0.2cm}

We first present the ABC protocol~\cite{Zhongjie-NDSS-2018} and the protection methods implemented in this protocol. We then explain in detail the impersonation attack scheme~\cite{benegui2020breach} that is able to bypass the ABC protocol. Finally, we describe our multi-modal ABC protocol that can detect the impersonation attack.

%\vspace{-0.2cm}
\subsection{ABC Protocol}
%\vspace{-0.1cm}

The protocol defined by Zhongje et al.~\cite{Zhongjie-NDSS-2018} is represented by a two stage authentication system, described as follows. The first stage of the protocol is represented by the registration phase, in which the user's smartphone is enrolled in the system using an image, denoted as $I_{(r)}$, taken using the built-in camera.  The server (verifier) employed in the protocol computes an estimate of the camera PRNU fingerprint $\hat{K}_{(c)}$ from the received image, creating a device profile used for authentication purposes.

In the second stage of the protocol, namely the authentication phase, three sequential actions are executed, as follows: $(i)$ the server generates two QR code images and presents them to the user, $(ii)$ the user takes a picture of each QR code and $(iii)$ sends the photos back to the server for verification. Each QR code contains embedded information representing the current transaction in progress. 
In step $(i)$, along with the embedded metadata, the QR code images produced by the server, contain a probe signal $\Gamma_{i}$ represented by a white Gaussian noise:

\begin{equation}\label{eq_auth_step1}
I_{i(s)} = QR(str_i, T_i) + \Gamma_{i}, \forall i \in \{1,2\}.
\end{equation}
In step $(ii)$, the user takes photos of the prompted images on a screen, using a preregistered device. The resulting photos, $I_{1(c)}$ and $I_{2(c)}$, are sent to the server for verification and user identification. The captured images, denoted by $I_{i(c)}$, should contain a noise residue $W_{i(c)}$ composed of the PRNU fingerprint of the user and the probe signal $\Gamma_i$:
\begin{equation}\label{eq_auth_step2}
I_{i(c)} = QR(str_i, T_i) + W_{i(c)}, \forall i \in \{1,2\},
\end{equation}
where the noise residue is formally defined as follows:
\begin{equation}\label{eq_noise_residue}
W_{i(c)} = \Gamma_{i} + {K}_{(c)}.
\end{equation}

In the last authentication step $(iii)$, the server performs a multi-step user validation and identification, such as QR code integrity check, camera fingerprint verification, forgery detection and probe signal verification, as detailed in~\cite{Zhongjie-NDSS-2018}. If all integrity checks pass successfully, the user is authenticated by the protocol, otherwise, the system will reject the transaction.

%\vspace{-0.2cm}
\subsection{ABC Protocol Defense Systems}
%\vspace{-0.1cm}

\subsubsection{Forgery detection:}

Zhongjie et al.~\cite{Zhongjie-NDSS-2018} propose an anti-forgery system that is able to protect the protocol from attackers that use counterfeit images, in the authentication phase. If an attacker tries to impersonate a victim, his fingerprint ${K}_{(a)}$ can be present in the forged image. Therefore, the protocol computes the noise residue $W_{i(c)}$ for each of the two received images and compares $W_{1(c)}$ and $W_{2(c)}$, according to the following equation:
\begin{equation}\label{eq_sim_noise}
PCE\left(W_{1(c)}, W_{2(c)}\right),
\end{equation}
where $PCE$ is the Peak-to-Correlation Energy~\cite{Goljan-IWDW-2008}. Furthermore, the PRNU fingerprint computed during user registration is also compared with the noise residue $W_{1(c)}$ extracted from $I_{1(s)}$, the similarity value being given by:
\begin{equation}\label{eq_sim_normal}
PCE\left(W_{1(c)}, \hat{K}_{(c)}\right).
\end{equation}

In case of forged images, the similarity between $W_{1(c)}$ and $W_{2(c)}$ is higher in comparison with the similarity of the noise residue $W_{1(c)}$ and the registered PRNU fingerprint $\hat{K}_{(c)}$. Hence, the ABC protocol uses the following equation to determine whether the images are forged or not:

\begin{equation}\label{eq_forgery_detection}
PCE\left(W_{1(c)}, W_{2(c)}\right) > PCE\left(W_{1(c)}, \hat{K}_{(c)}\right) + t,
\end{equation}
where $t$ is threshold that eliminates matching by chance due to noise.

An adversary can compute his own fingerprint ${K}_{(a)}$ and remove it from the forged images in order to fool the forgery detection system. To this end, Zhongjie et al~\cite{Zhongjie-NDSS-2018} proposed a removal detection system (described below) based on the probe signal embedded in the QR code images. 

\subsubsection{Removal detection:}

During an impersonation attack, the adversary has the aim to remove his fingerprint from the images sent to the verifier. Since the PRNU fingerprint is very similar, in terms of magnitude, to the probe signal embedded in the QR code images, removal of the fingerprint will inherently lead to the removal of the probe signal as well. Thus, when the verifier analyzes the images, the unique pattern noise $\Gamma_i$ will not be present. Therefore, the removal detection system will reject the attack. However, Zhonhjie et al.~\cite{Zhongjie-NDSS-2018} follow the assumption that an adversary computes his fingerprint using the photos captured during the authentication phase, in which the probe signal is embedded. Contrary to their assumption, as we are about to detail further, an adversary can compute his fingerprint from any photo resulted from the device used in the attack. In consequence, we exploit this vulnerability in our attack described below.

%\vspace{-0.2cm}
\subsection{An Attack for the ABC Protocol}
%\vspace{-0.1cm}

Zhongjie et al.~\cite{Zhongjie-NDSS-2018} assumed in their work that a potential adversary computes the PRNU fingerprint using photos captured during the authentication phase. We propose an attack that exploits a vulnerability in the assumption of Zhongjie et al.~\cite{Zhongjie-NDSS-2018}, namely that an attacker can compute the camera fingerprint of the device used in the impersonation attack, $\hat{K}_{(a)}$, at any given time prior to the attack. 
Thus, we further %describe in our experiments presented Section~\ref{sec_Experiments}, 
consider the case in which a pre-computed fingerprint is used within the impersonation attack. % gaining access to give photos from both entities involved in the
% using one or five images
In addition, considering that an adversary can gain access to one or a few photos posted on social media platforms by a potential victim, an impersonation attack becomes feasible. In the experiments presented in Section~\ref{sec_Experiments}, we consider two attack scenarios. The first one uses only one photo to estimate the victim's PRNU fingerprint, following the same setup considered by Zhongjie et al.~\cite{Zhongjie-NDSS-2018}. The second one uses five photos to estimate the victim's PRNU fingerprint, showing that the success rate of our attack can be improved further.
% Moreover, we also experiment using one image with the aim to compare the results obtained by Zhongjie et al.~\cite{Zhongjie-NDSS-2018}. 
% The empirical results demonstrate that more images used in the attack yield better success rates.

Our attack works as follows.
In the authentication phase, the attacker uses the built-in camera to capture the two photos generated by the server with the aim of forging and sending them back for verification. The photos taken by the attacker are defined as follows:
\begin{equation}\label{eq_auth_step2_attack}
I_{i(c)} = QR(str_i, T_i) + \Gamma_{i} + {K}_{(a)}, \forall i \in \{1,2\}.
\end{equation}
Before sending them to the verifier, the images contain the PRNU fingerprint ${K}_{(a)}$ of the adversary instead of the fingerprint ${K}_{(c)}$ of the victim. As described by Zhongjie et al.~\cite{Zhongjie-NDSS-2018}, if the adversary tries to remove his fingerprint $\hat{K}_{(a)}$ using the images $I_{i(c)}$ defined in Equation~\eqref{eq_auth_step2_attack} which include the probe signal $\Gamma_{i}$, the attack would be stopped by the ABC Removal Detection system. However, in our approach, the adversary fingerprint $\hat{K}_{(a)}$ is pre-computed using other photos, different from the ones involved in the authentication phase. Hence, the probe signal $\Gamma_{i}$ is only slightly altered, but not entirely removed. Furthermore, adding the pre-computed victim's fingerprint results in a set of forged photos $I_{i(f)}$ that can bypass the defense systems of the ABC protocol. The forged images are defined as follows:
\begin{equation}
I_{i(f)} \approx QR(str_i, T_i) + \Gamma_{i} + \hat{K}_{(c)}, \forall i \in \{1,2\}.
\end{equation}
When the verifier employs $(i)$ Forgery Detection to determine if the images are forged and $(ii)$ Removal Detection to assess if the probe signal was removed, the system will find that $\Gamma_{i}$ is included in the received images and the PRNU fingerprint present in the images is the one of the impersonated person. Therefore, the proposed attack scheme bypasses both systems of the ABC protocol. However, due to approximation errors involved in the attack process, the system is able to block about one in every two attacks, as detailed in our experiments below. Still, the attack success rate leaves the standard ABC protocol vulnerable to our attack scheme. Further details about our attack scheme are provided in our preliminary work~\cite{benegui2020breach}.

\begin{figure}[!t]
\begin{center}
\includegraphics[width=1.0\linewidth]{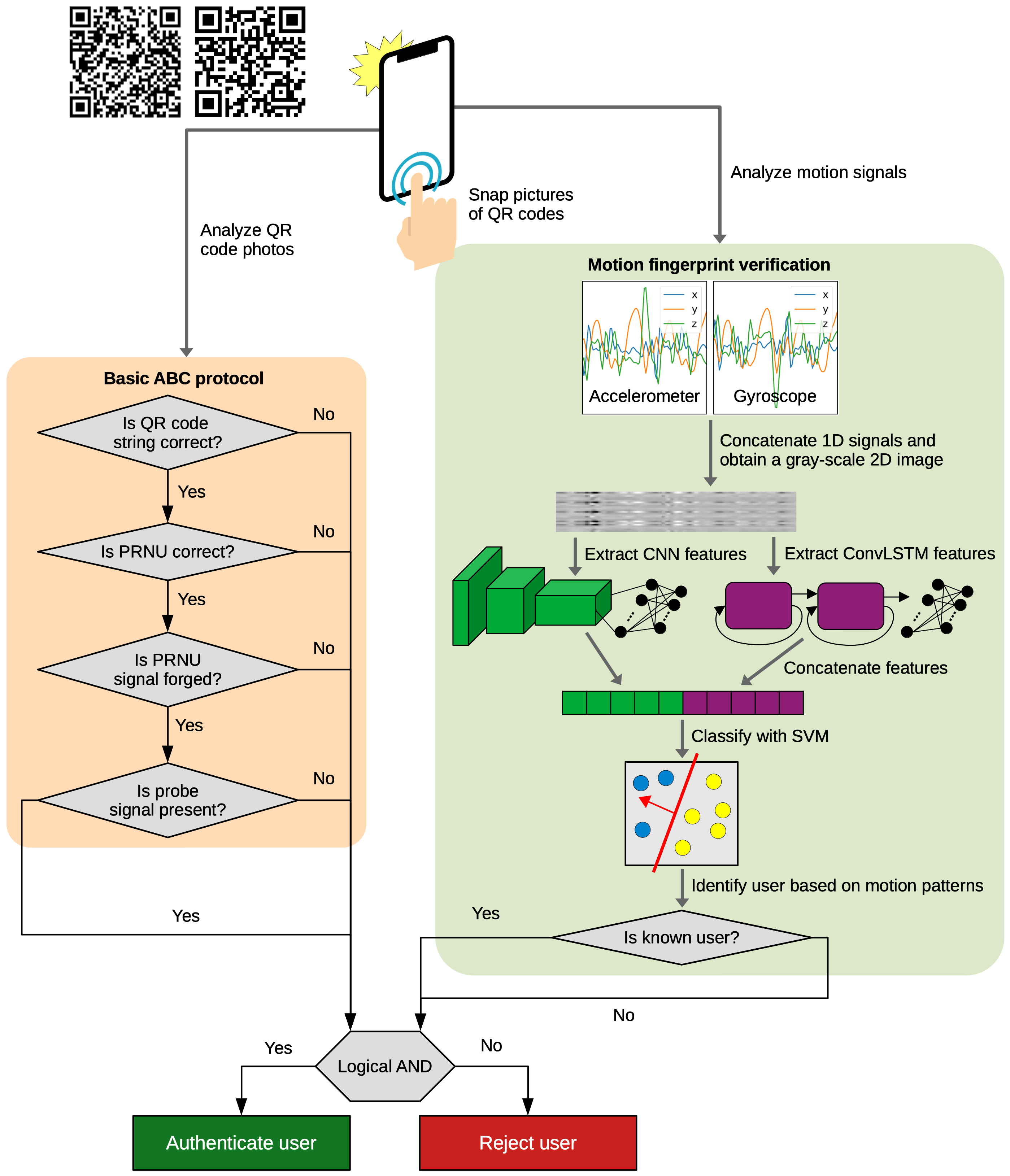}
\end{center}
\vspace*{-0.3cm}
\caption{An overview of the proposed multi-modal ABC protocol. The basic ABC protocol is augmented with a deep learning system that analyzes the motion fingerprint. Best viewed in color.}
\label{fig_pipeline}
\end{figure}

\subsection{Proposed Multi-Modal ABC Protocol}

As noted above, the ABC protocol is vulnerable, being possible to bypass both protection systems of the protocol. We further propose an extension of the protocol which incorporates motion sensor data, as an additional authentication factor that restores the overall protection level of the system. Unlike photos, which are commonly posted on social media platforms, motion sensor signals are not openly posted by smartphone users on the Internet, mainly due to the lack of interest to do such such thing. Therefore, our multi-modal authentication protocol is equipped with an enhanced security mechanism that cannot be exploited by potential attackers. Another advantage of our multi-modal protocol is that the introduction of the second modality does not burden the user with additional steps during authentication, i.e.~the second authentication factor is implicit. Our novel protocol is illustrated in Figure~\ref{fig_pipeline}. The stages involved in our multi-modal ABC protocol are described next.

%Benegui et al.~\cite{Benegui-Access-2020} demonstrate that user identification based on motion sensors using a pre-trained 6-layers CNN as a feature extractor for a binary SVM can attain a top accuracy of $96.72\%$ with a FAR of $3.10\%$ and a FRR of $3.45\%$. Combined with the ABC protocol, our proposed extension reduces the false acceptance ratio to as less as $0.07\%$, as we further demonstrate in our experiments.

\subsubsection{Motion Signal Recording}

In the authentication phase, we collect three-axis motion signals from two sensors, the gyroscope and the accelerometer, while the user is taking photos of the QR codes. A motion sample is formed of six discrete signals. There are three signals for each sensor, such that each signal corresponds to one of the three spatial axes $(x, y, z)$. The signals are continuously recorded at 100 Hz during the whole authentication process, but we only keep the signals recorded within a window of 1.5 seconds, starting with 0.5 seconds before the user touches the button that causes the smartphone to take the first photo $I_{1(c)}$ required for verification. Since each signal is recorded at 100 Hz for 1.5 seconds, it is composed of roughly 150 values. The recorded signals are associated with the image $I_{1(c)}$, being sent together to the sever for verification. 

\subsubsection{Motion Signal Pre-processing}

On the sever side, the collected discrete signals are processed in order to turn them into mono-channel images, enabling us to employ CNNs and ConvLSTMs to learn neural embeddings. Even if, in theory, the motion signals should be recorded at exactly 100 Hz, we observed that, in practice, mobile operating systems (iOS and Android) will not report precisely 100 values per second at perfectly equal time intervals, likely being influenced by the different processes running on each mobile device. Moreover, the accelerometer and gyroscope generate decoupled motion events at a frequency that is close to 100 Hz, but not exactly equal to 100 Hz. Hence, the signals generated by the two independent motion sensors are not necessarily of the same length. To this end, we must normalize all motion signals to a fixed length. As the signals are recorded for 1.5 seconds at 100 Hz, we start with the assumption that each signal should be formed of exactly 150 discrete values. Thus, we resize the overlength or underlength signals to a fixed length of 150 values through linear interpolation. After resizing a signal, we subtract its minimum magnitude to eliminate negative values. Since the signal corresponding to one axis can have a different magnitude scale than the signals corresponding to other axes, e.g.~when the magnitude of the motion along one axis is much higher than the magnitude of the motion along another axis, we rescale each discrete signal using the $L_2$-norm. Finally, we resample all values to a range between $0$ and $1$. %, in order to use the full range of values available for gray-scale images.

\subsubsection{Learning Neural Embeddings}

To learn discriminative neural embeddings from the motion signals, we consider two neural network architectures, a CNN and a ConvLSTM, which we train on a multi-way user classification task on the HMOG data set~\cite{Sitova-TIFS-2016}. The learning stage is an offline step that needs to be carried out before deploying the multi-modal ABC protocol in testing or production environments.

Our CNN model is composed of three convolutional (conv) layers, followed by two fully-connected (fc) layers and one Softmax classification layer, having a total of six layers. We set the number of filters in the first conv layer to 32. Following AlexNet~\cite{Krizhevsky-NIPS-2012} and VGG~\cite{Simonyan-ICLR-14} architectures, the conv layers get wider as they move farther away from the input. This is because conv layers closer to the input learn low-level features, e.g.~edges or corners, which are generally useful for all object classes. Conv layers closer to the output learn highl-level features, which are specialized to particular object classes. Hence, our second conv layer is composed of 64 filters, while the third conv layer is composed of 128 filters. Following recent CNN architectures such as ResNet~\cite{He-CVPR-2016}, we employ conv filters with a small receptive field of $3\times3$ components. The filters are applied at a stride of 1. All activation maps are zero-padded to preserve the spatial dimension. We placed a max-pooling layer after each conv layer. The max-pooling layers have a pool size of $2\times2$ and are applied at stride $2$. Each fc layer is formed of 256 units with dropout~\cite{Srivastava-JMLR-2014} at a rate of $0.4$, to prevent overfitting. All layers have Rectified Linear Units (ReLU)~\cite{Nair-ICML-2010} activations, except for the classification layer. As our neural models must solve a multi-class classification problem, we employ Softmax activation in the last layer, such that the final output provides the probability for each class. We underline that the last layer is formed of 50 neurons, this being the number of classes in our data set. We train the neural networks with the Adam optimizer \cite{Kingma-ICLR-2015}, minimizing the categorical cross-entropy loss.

Similar in size to the proposed CNN, the architecture of the ConvLSTM is composed of six layers. The first layer of the network is a convolutional LSTM layer containing 64 kernels. The first layer is followed by a second conv LSTM layer with 128 filters and a third conv LSTM layer with 256 filters. All filters have a spatial support of $1\times3$. We underline that a very common practice is to design LSTM architectures having no more than two or three recurrent layers, which are more than enough to capture the temporal aspects of the input signals. Thus, we use only three conv LSTM layers, flattening the activation maps resulting after the last recurrent layer. Then, we have two fc layers, each with 256 neurons. All conv LSTM and fc layers are equipped with ReLU~\cite{Nair-ICML-2010} activations. The sixth and last layer provides the final class probabilities, being composed of 50 neurons with Softmax activation (each neuron provides the probability for one user in the multi-class user classification data set). As for our CNN model, we employ the Adam optimizer~\cite{Kingma-ICLR-2015} to minimize the categorical cross-entropy loss.

We underline that the CNN and the ConvLSTM are trained prior to their integration in our multi-modal ABC protocol. After training the CNN and the ConvLSTM models on the multi-class user classification task, we remove the Softmax layer from each architecture, thus using the feature vectors from the last fc layer. Since the fc layers have 256 neurons each, we obtain 256-dimensional feature vectors from each model. Our final neural embeddings are 512-dimensional feature vectors obtained by concatenating the corresponding CNN and the ConvLSTM feature vectors. In the experiments, we show the benefit of combining the CNN and the ConvLSTM embeddings.

% and forwarded as inputs for two deep neural networks, a 6-layers CNN and a 6-layers ConvLSTM network, pretrained for few-shot user identification. The embeddings resulted, are further passed as inputs to an SVM which performs user identification. 

% We conduct the experiments based on research done by Benegui et al. \cite{Benegui-Access-2020}. We select the best performing architecture, specifically the 6-layers architecture, for our deep neural networks used as feature extractors. In our experiments, we remove the output layers, namely Softmax, from both architectures and extract embeddings (feature vectors) from the last fully-connected layer. The resulting embeddings have a length of 256 components, equal to the number of neurons of the layer. The complete configuration of both networks is provided in \cite{Benegui-Access-2020}.

\subsubsection{Motion Sensor Fingerprints}

For the smartphone identification problem based on motion sensors, we utilize a binary SVM classifier, which receives as input the neural embeddings resulted from our pre-trained deep neural networks. We trained an SVM for each smartphone device, using the neural embeddings computed during the registration phase of the respective device as positive examples. Since the SVM requires negative examples as well, we use a pool of negative examples that is independent of all smartphone devices considered in our experiments. This ensures that the SVM models are never trained on examples belonging to attackers, which would lead to unrealistically high accuracy levels.

% To further enhance the results obtained, we employ concatenation of embeddings from both neural networks into a single feature vector of 512 components. The embeddings concatenation produces the best results, as detailed in the following experiments.
 
Formulating our smartphone identification task as a binary classification problem, the SVM learns a linear discriminant function $f$ that outputs the label $+1$ for an input neural embedding belonging to the registered smartphone and the label $-1$ for a neural embedding belonging to an adversarial device. The linear function $f$, can be expressed as follows:
\begin{equation}\label{eq_binary_func}
\begin{split}
f(x)=\mbox{sign}(\langle w, x \rangle + b),
\end{split}
\end{equation}
where $x$ represents a feature vector, $w$ and $b$ denote the vector of weights and the bias term of the classifier, and $\langle \cdot, \cdot \rangle$ represents the scalar product. As explained earlier, the feature vector $x$ contains 512 values and is generated by concatenating 256-dimensional neural embeddings from the CNN and the ConvLSTM models, respectively. 

An SVM classifier~\cite{Cortes-ML-1995} computes the parameters $w$ and $b$ that represent the hyperplane which divides the training examples into two classes with maximum margin. Formally, the SVM model finds the parameters $w$ and $b$ that satisfy the optimization criterion defined below:
\begin{equation}\label{eq_objective}
\begin{split}
\min_{w,b}\frac{1}{n}\sum\limits_{i=1}^n[1-y_i(\langle w,x_i \rangle + b)]_+ + C \lVert w \rVert^2 ,
\end{split}
\end{equation}
where $n$ is the number of examples from the training set, $y_i$ is the label ($+1$ or $-1$) associated to the training example $x_i$, $C$ represents a regularization hyperparameter, $[x]_+=\max \lbrace x, 0 \rbrace$ and $\lVert \cdot \rVert^2$ is the $L_2$-norm.

Let $f^j$ be the discriminant function corresponding to a device $j$. We interpret the corresponding weights $w^j$ and the bias $b^j$ as a motion sensor fingerprint of the smartphone device $j$. Upon learning the parameters $w^j$ and $b^j$ by optimizing Equation~\eqref{eq_objective}, during authentication, we just need to apply the following equation on a neural embedding $x$ from an unknown a device in order to identify the respective device as device $j$:
\begin{equation}\label{eq_binary_func_2}
\begin{split}
f^j(x)=\mbox{sign}(\langle w^j, x \rangle + b^j).
\end{split}
\end{equation}
%Furthermore, during the registration phase, motion sensor values are processed, passed through the deep neural networks for feature extraction, the result being used to train a binary classifier, namely SVM~\cite{Cortes-ML-1995}, for the user identification task. Thus, during the authentication phase, additionally to the images processed by the ABC protocol, motion sensor data associated with the authentication samples are sent as input to the binary SVM with the aim to identify if the user is a legitimate user or belongs to a non-legitimate class. Hence, the enhanced ABC protocol has another layer of security through the classifier. 
We note that the function $f^j$ is applied in conjunction with the ABC protocol. 
%After both the ABC systems and the binary classifier offer an output for the sample provided, the results are merged. 
If the multi-modal sample (composed of two photos and a set of motion sensor signals) collected during an authentication session passes both camera and motion sensor verification layers, then the corresponding smartphone is identified as a legitimate smartphone and authorization takes place. This means that if a sample does not pass one of the verification layers, then the authentication is rejected. Empirical evidence shows that by employing our additional layer of security based on motion sensor signals, we can restore the security level of the multi-modal ABC protocol.

\section{Experiments}
\label{sec_Experiments}
%\vspace{-0.1cm}
\subsection{Data Sets}
%\vspace{-0.1cm}

To evaluate the success rate of our attack on the standard ABC protocol and on the proposed multi-modal ABC protocol, we collected a multi-modal data set consisting of images and motion signals. 

Using six different smartphone devices, we composed a data set consisting of 105 images per device along with the motion sensor data generated during the photo capture session. Based on previous works on PRNU estimation~\cite{Quiring-WIFS-2015,Zhongjie-NDSS-2018}, we extracted sub-images of $1000\!\times\!750$ pixels, starting from the top left corner, to compute PRNU fingerprints. 

For our enhanced ABC protocol, we recorded motion signals for $0.5$ seconds before and $1$ second after pressing the camera shutter. The accelerometer and gyroscope signals, each represented on three axes, are recorded at $100$ Hz, resulting in signals of $150$ discrete values in the time domain. Each photo in our data set is thus associated with a multi-dimensional motion signal of $6 \times 150$ values, where $6$ is the number of motion sensors (accelerometer, gyroscope) multiplied by the number of axes ($x$, $y$, $z$).
To allow others to reproduce our results, we will provide our data set for non-commercial use to those who send their request by mail to one of the authors.

Due to the small set of motion signals in our multi-modal data set, we pre-train our CNN and ConvLSTM models on a subset of 50 users from the HMOG data set~\cite{Sitova-TIFS-2016}. Following the experimental settings described in~\cite{Benegui-Access-2020}, we collect motion signals for 200 tap events per user, generating a data set of 10.000 data samples in total. Employing an $80\%$-$20\%$ split of the data, we utilize 8.000 samples for training and 2.000 for validation.

\subsection{Organization of Experiments}

Considering our ABC attack scheme~\cite{benegui2020breach} as well as the original attack scheme proposed by Zhongjie et al.~\cite{Zhongjie-NDSS-2018}, our first set of experiments aims to test the security level of the ABC protocol. We hereby consider two scenarios. 

In the first scenario, we use five images to compute the PRNU fingerprint of each device and 100 images to simulate authentications of a registered user. Furthermore, we use the same set of 100 images to perform simulated impersonation attacks on the other devices. Therefore, we perform 600 valid authentications and 3000 attacks. This scenario is motivated by the fact that attackers can often obtain more than one image (in our scenario, we consider five images) from social media posts to compute a victim's PRNU fingerprint. Naturally, the attacker is free to use as many photos as necessary (we limit ourselves to five images) to compute his own PRNU fingerprint prior to the attacks.

In the second scenario, the experiments are conducted using one image for PRNU fingerprint estimation for both victims and attackers, precisely following the setting described by Zhongjie et al.~\cite{Zhongjie-NDSS-2018}. Here, our aim is to demonstrate that our attack scheme can bypass the ABC protocol, without any change with respect to the introductory work~\cite{Zhongjie-NDSS-2018}. The only difference with respect to our first scenario is the number of images used for PRNU estimation in the registration phase. Thus, as in the first scenario, we utilize the same number of devices and images during authentication, resulting in 600 valid sessions and 3000 impersonation attacks.

To test our enhanced ABC protocol, our second and last set of experiments employs deep learning models for user classification based on motion sensors. Upon training the deep learning models in the same setting as in our previous work~\cite{Benegui-Access-2020}, our aim is to test the efficiency of the proposed attack scheme~\cite{benegui2020breach} on our multi-modal data set. Noting that we need to fit a machine learning model on several (more than one) motion signals, we employ the same setting as in the first scenario, considering five images (and associated motion signals) to compute the PRNU (and motion sensor) fingerprints. This limits our multi-modal ABC protocol to use a mandatory lower bound (five) on the number of registration sessions. Nonetheless, this limitation fades away in front of the benefit, namely resistance to impersonation attacks.

%\vspace{-0.2cm}
\subsection{Evaluation Details}
%\vspace{-0.1cm}

\subsubsection{Evaluation Measures} 

We report the number of successful attacks (false acceptances) as well as the \emph{false acceptance rate} (FAR), which is typically defined as the ratio of the number of false acceptances divided by the number of authentication attempts. A \emph{false acceptance} is an instance of a security system, in our case the original or the multi-modal ABC protocols, incorrectly verifying an unauthorized person, e.g.~an impersonator. We underline that our attack does not impact the \emph{false rejection rate} (FRR) of the ABC protocol, i.e.~the FRR is similar to that reported in~\cite{Zhongjie-NDSS-2018}. For the multi-class user classification experiments on HMOG, we report the classification accuracy rate. For the multi-modal ABC protocol, we report the accuracy, FAR and FRR values, respectively.

%\vspace{-0.4cm}
\subsubsection{Evaluation Protocol}

The main goal of the first two sets of experiments is to validate the attack scheme proposed in~\cite{benegui2020breach}. While reporting the FAR values for our attack is necessary, we also have to validate that the forgery detection (FD) system and the removal detection (RD) system of the ABC protocol work properly. For this reason, we need to perform attacks as described in~\cite{Zhongjie-NDSS-2018}. Our aim is to show that the protection systems of the ABC protocol are indeed able to reject the attacks specified in~\cite{Zhongjie-NDSS-2018}, while not being able to detect our own attack. 

When testing the ABC protocol or the multi-modal ABC protocol against attacks, each of the $n$ smartphone devices takes turn in being considered as the victim's device. In order to perform attacks, the remaining $n-1$ devices are considered to belong to adversaries. Each adversary performs 100 attacks. Given that our data set consists of $n=6$ devices, we obtain a number of 3000 ($6 \times 5 \times 100$) attacks. For each attack, we determine if it passes undetected by the Forgery Detection system and by the Removal Detection system. We consider a successful attack only if it succeeds to cross both Forgery Detection and Removal Detection systems.
We count the number of successful attacks and compute the corresponding FAR at different PCE thresholds between 10000 and 50000, using a step of 100. We note that the threshold values are generally higher than those used in~\cite{Zhongjie-NDSS-2018}, because we compute the PRNU fingerprints on larger images. We determine the \emph{optimal threshold} as the threshold that provides a FAR of roughly $0.5\%$ for the attack scheme detailed in~\cite{Zhongjie-NDSS-2018}, because Zhongjie et al.~\cite{Zhongjie-NDSS-2018} report a FAR of $0.5\%$ in their paper. We note that they selected the threshold that corresponds to equal FAR and FRR.

%\vspace{-0.2cm}
\subsection{Attacking the ABC Protocol}

\subsubsection{Results with Five Images for PRNU Estimation}
%\vspace{-0.1cm}

\begin{table}[!th]
\setlength\tabcolsep{2.2pt}
\renewcommand{\arraystretch}{1.2} 
\caption{False acceptance rates (FAR) and number of successful attempts (in parentheses) for the attack scheme proposed in~\cite{benegui2020breach} versus the attack scheme detailed in~\cite{Zhongjie-NDSS-2018}, when five images are used for PRNU estimation in the standard ABC protocol. False acceptance rates are computed for five PCE thresholds between 10000 and 50000. Results (highlighted in bold) for the optimal PCE threshold (22500) are also included. For each attack scheme, we report the false acceptance rates for the Forgery Detection (FD) system, the Removal Detection (RD) system, and both (FD+RD).}
\small{
\begin{center}
\begin{tabular}{|c|c|c|c|c|c|c|}
\hline
\textbf{Threshold} & 
\multicolumn{3}{| c |}{\textbf{Our attack~\cite{benegui2020breach}}}  &
\multicolumn{3}{| c |}{\textbf{ABC attack~\cite{Zhongjie-NDSS-2018}}} \\
\cline{2-7}
        & \textbf{FD FAR}    & \textbf{RD FAR}    & \textbf{FD+RD FAR}     & \textbf{FD FAR}    & \textbf{RD FAR}    & \textbf{FD+RD FAR}\\
%        & FAR       & FAR       & FAR           & FAR       & FAR       & FAR \\
\hline
\hline
10000   & $74.5\%$ (2235)  & $90.0\%$ (2701)  & $70.7\%$ (2122)  & $87.6\%$ (2628) & $30.7\%$ (920) & $25.9\%$ (776) \\
\hline
20000   & $62.6\%$ (1879)  & $83.9\%$ (2517) & $57.5\%$ (1726)   & $82.9\%$ (2487)  & $2.4\%$ (73)  & $1.5\%$ (45) \\
\hline
{\bf 22500} &$\mathbf{60.0\%}$ (1800)  &$\mathbf{81.7\%}$ (2451)  &$\mathbf{54.1\%}$ (1624)  & $\mathbf{81.5\%}$ (2446) & $\mathbf{1.0\%}$ (31) &$\mathbf{0.5\%}$ (16) \\
\hline
30000   & $53.3\%$ (1600)  & $76.4\%$ (2292) & $45.9\%$ (1378)     & $78.3\%$ (2350)  & $0.0\%$ (0)  & $0.0\%$ (0) \\
\hline
40000   & $48.1\%$ (1444) & $72.8\%$ (2184)  & $40.6\%$ (1219)     & $74.5\%$ (2234)  & $0.0\%$ (0)   & $0.0\%$ (0) \\
\hline
50000   & $43.8\%$ (1315)  & $69.7\%$ (2090)  & $35.7\%$ (1071)      & $71.7\%$ (2150)  & $0.0\%$ (0)   & $0.0\%$ (0) \\
\hline
\end{tabular}
\end{center}
\label{tab_results_far}
}
\end{table}

\begin{figure}[!t]
\begin{center}
\includegraphics[width=0.9\linewidth]{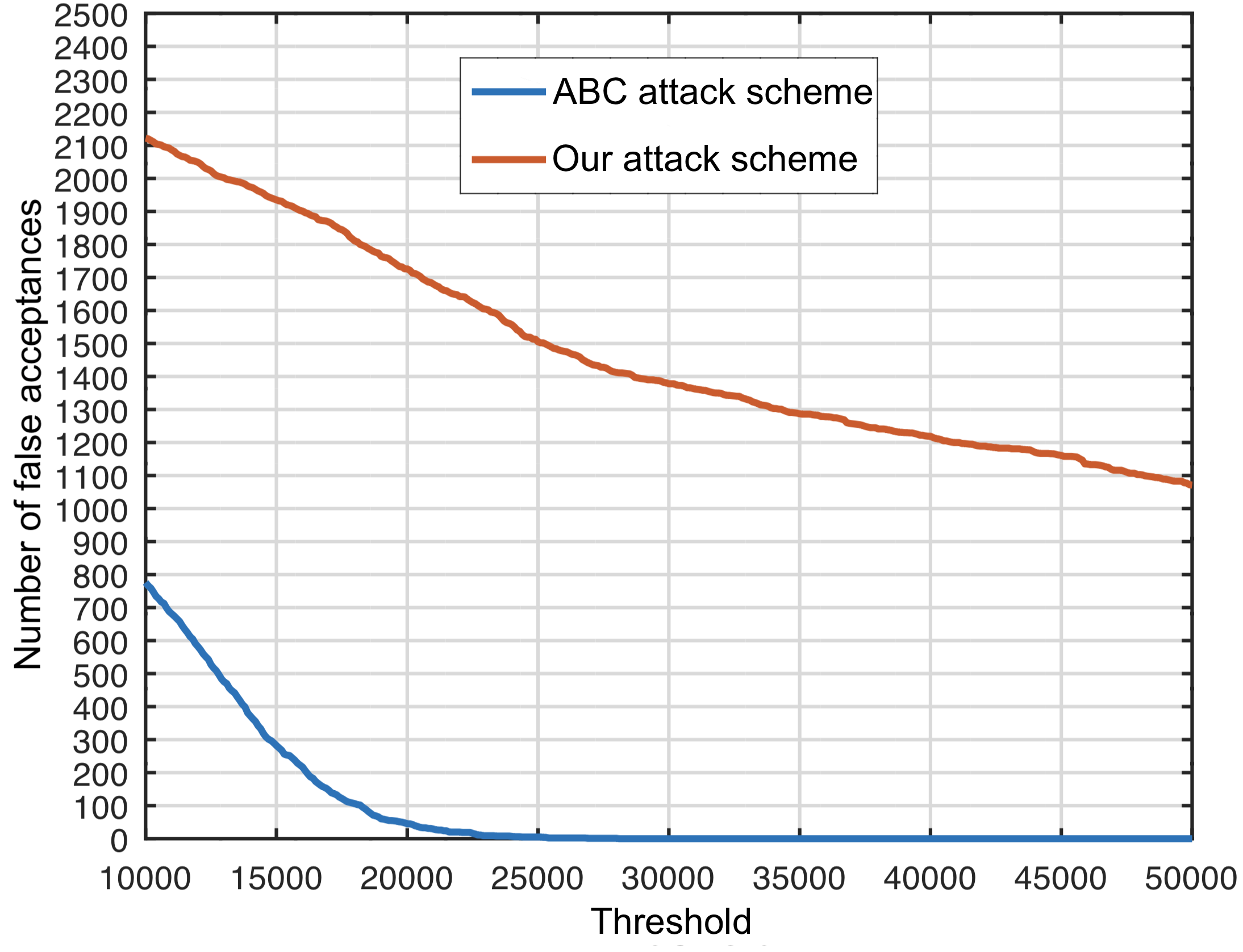}
\end{center}
\vspace*{-0.3cm}
\caption{Number of false acceptances (on the vertical axis) bypassing both Forgery Detection and Removal Detection systems of the original ABC protocol, for the attack scheme proposed in~\cite{benegui2020breach} versus the attack scheme detailed in~\cite{Zhongjie-NDSS-2018}, when five images are used for PRNU estimation. False acceptances are counted for multiple PCE thresholds (on the horizontal axis) between 10000 and 50000, with a step of 100. Best viewed in color.}
\label{fig1_step100}
\end{figure}

In Figure~\ref{fig1_step100}, we show the number of false acceptances generated by our attack scheme~\cite{benegui2020breach} in comparison with the attack described by Zhongjie et al.~\cite{Zhongjie-NDSS-2018}. Both attacks employ five images for fingerprint estimation, results being compared using multiple PCE thresholds between 10000 and 50000. Attacks that bypass the ABC protocol defense systems, such as Forgery Detection and Removal Detection, are counted as valid authentications. Our attack scheme, obtains a FAR of $0.5\%$ at a threshold of 22500, similar to what Zhongjie et al.~\cite{Zhongjie-NDSS-2018} obtained in their experiments. Thus, we consider this value as being the optimal threshold. However, we emphasize that for all thresholds between 10000 and 50000, our attack scheme is significantly more successful.

Table~\ref{tab_results_far} compares our attack scheme to the attack scheme of Zhongjie et al.~\cite{Zhongjie-NDSS-2018}, for different thresholds between 10000 and 50000. At the selected optimal threshold of 22500, our attack scheme achieves a FAR equal to $54.1\%$ (1624 successful attacks) while the ABC protocol attack scheme described in~\cite{Zhongjie-NDSS-2018} achieves a FAR of $0.5\%$. To further prove that the results are consistent with the numbers reported in~\cite{Zhongjie-NDSS-2018}, we compute the False Rejection Rate (FRR) by doing 600 authentications with registered devices, obtaining a similar FRR (under $0.1\%$). Our FAR and FRR show that half of the attacks are successful. Therefore, the ABC protocol is unsafe when an attacker has access to a victim's photos.

%Our attack scheme bypasses the Removal Detection system with a much higher FAR compared to the original scheme considered in~\cite{Zhongjie-NDSS-2018}. 

While our attack can bypass the Removal Detection system with a much higher FAR than the attack scheme considered in~\cite{Zhongjie-NDSS-2018}, it gives slightly lower FAR values in trying to bypass the Forgery Detection system, because the attacker's PRNU fingerprint is computed on a different set of images than the two QR code images used during authentication. More specifically, the lower FAR rates are generated by the approximation errors between the PRNU estimation $\hat{K}_{(a)}$ and the actual PRNU fingerprint $K_{(a)}$ found in the QR code images. Nevertheless, our proposed attack generates higher false acceptance rates even when both systems are considered together. Therefore, the results presented in Table~\ref{tab_results_far} strongly indicate that our attack scheme is very powerful against the ABC protocol, succeeding in one of every two attacks. 

\subsubsection{Results with One Image for PRNU Estimation}

\begin{figure}[!t]
\begin{center}
\includegraphics[width=0.9\linewidth]{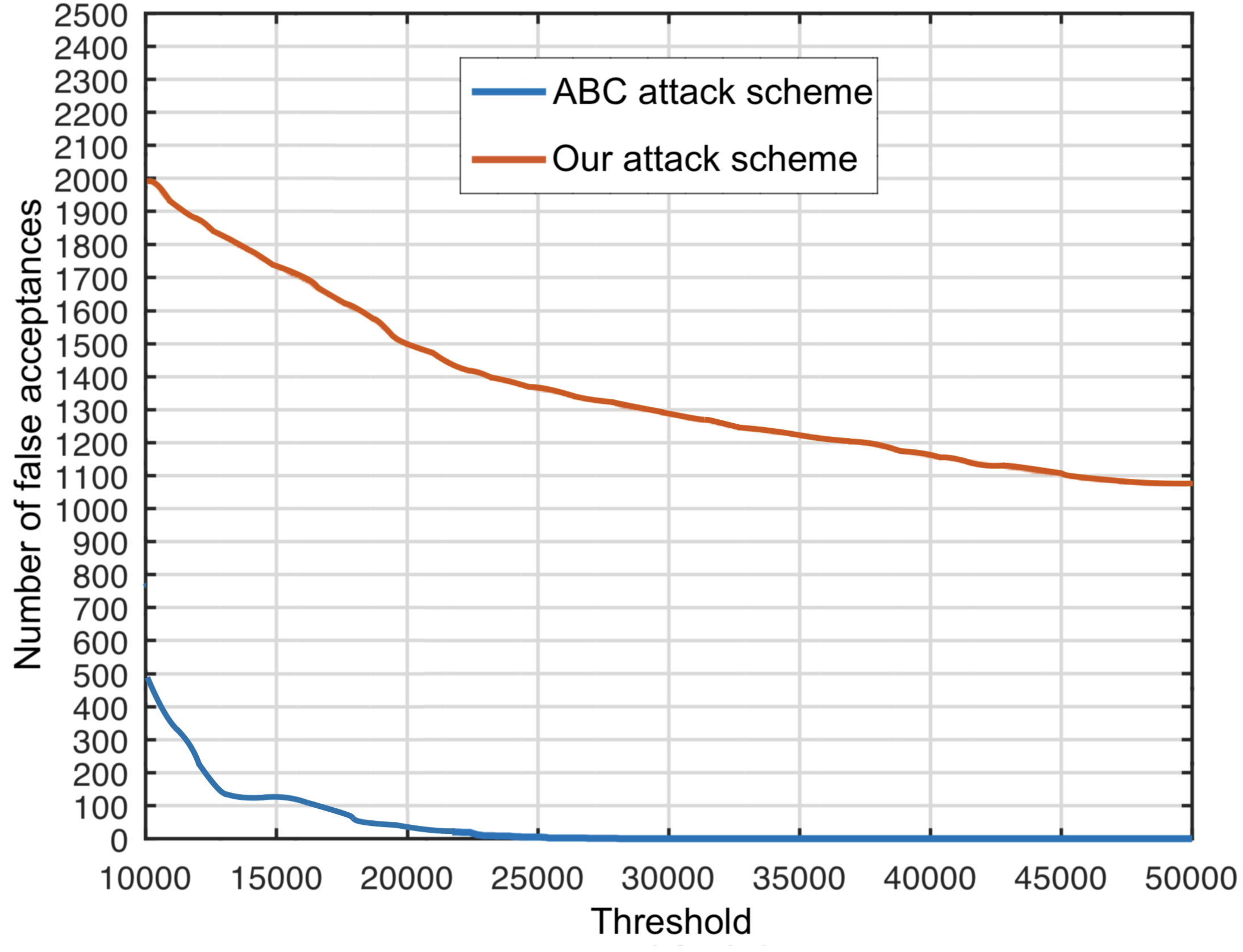}
\end{center}
\vspace{-0.3cm}
\caption{Number of false acceptances (on the vertical axis) bypassing both Forgery Detection and Removal Detection systems of the original ABC protocol, for the attack scheme proposed in~\cite{benegui2020breach} versus the attack scheme proposed in~\cite{Zhongjie-NDSS-2018}, when one image is used for PRNU estimation. False acceptances are counted for multiple PCE thresholds (on the horizontal axis) between 10000 and 50000, with a step of 100. Best viewed in color.}
\label{fig2_step100}
\end{figure}

By using only one image for PRNU estimation and the same PCE thresholds between 10000 and 50000, Figure~\ref{fig2_step100} shows the number of false acceptances generated by our proposed attack scheme in comparison to the attack scheme considered by Zhongjie et al.~\cite{Zhongjie-NDSS-2018}. With only one image to compute both the adversary's and the victim's PRNU fingerprints, this setting is more difficult and less likely to succeed. We carry this experiment to have an apples-to-apples comparison to Zhongjie et al.~\cite{Zhongjie-NDSS-2018}. 

Comparing Figure~\ref{fig1_step100} and Figure~\ref{fig2_step100}, we can draw two conclusions. First, the number of attacks that bypass the ABC protocol defense systems is lower when utilizing one image compared to the setting in which we use five images. Second, we can see that even in the harder setting, the number of successful attacks is large. Therefore, our attack scheme still represents a great threat for the ABC protocol. Considering the optimal threshold value of 22500, the number of successful attacks is 1430 (representing a FAR of $47.7\%$). The empirical results show that the ABC protocol is still vulnerable, regardless of the total number of images considered for PRNU estimation. 

We thus conclude that the protection systems of the ABC protocol, namely the Forgery Detection and Removal Detection systems, do not suffice to prevent our attack scheme proposed in~\cite{benegui2020breach}. 

\subsection{Multi-way Classification Results with Deep Models}

\begin{table}[!t]
\renewcommand{\arraystretch}{1.2} 
\caption{Multi-way user classification accuracy rates on the HMOG data set, obtained by a baseline model based on handcrafted features versus our CNN and ConvLSTM models and their combination represented by an ensemble that merges the two models.}
\label{table_multiway}
\centering
\begin{tabular}{|l|c|}
\hline
\textbf{Method} & \textbf{Accuracy}\\
\hline
Shen et al.~\cite{Shen-Sensors-2016} & 87.30\% \\
\hline
CNN             & 96.37\% \\
\hline
ConvLSTM        & 96.18\% \\
\hline
CNN+ConvLSTM    & 96.74\% \\
\hline
\end{tabular}
\end{table}

In our multi-way user classification experiments on HMOG, we trained the CNN model on mini-batches of 32 samples. The model is optimized for a maximum 50 epochs using a learning rate of $10^{-3}$. However, in order to prevent overfitting, we applied early stopping at around 40 epochs. We employ the same hyperparameters for the ConvLSTM model, namely a learning rate of $10^{-3}$ and mini-batches of 32 samples. The ConvLSTM is also trained for a maximum of 50 epochs, but the optimization is halted after 20 epochs due to early stopping. We also consider an ensemble that employs an SVM as meta-leaner on top of concatenated CNN and ConvLSTM neural embeddings. The SVM uses $C=100$ for regularization and is based on the RBF kernel. As baseline, we add a model based on invariant handcrafted features, as proposed by Shen et al.~\cite{Shen-Sensors-2016}. The corresponding results are presented in Table~\ref{table_multiway}. We observe that the CNN model attains an accuracy of $96.37\%$, while the ConvLSTM performs slightly worse, attaining an accuracy of $96.18\%$. By merging the two deep models, we obtain superior results, attaining an accuracy of $96.74\%$. We underline that our final results are very good, considering that the baseline based on handcrafted features attains an accuracy of only $87.30\%$ on the 50-way user classification task.

\subsection{Attacking the Multi-Modal ABC Protocol}

We further present the empirical results obtained after using our attack scheme on the multi-modal ABC protocol, which combines PRNU and motion sensor fingerprints. We conducted the experiments using five samples to estimate the fingerprints during the attacks, irrespective of the modality. As shown earlier, this scenario is more difficult (the attack exhibits a higher penetration rate) than using a single image for fingerprint estimation (see Figures~\ref{fig1_step100} and~\ref{fig2_step100}).

% In our enhanced protocol, we employ motion sensor data in both authentication and registration phase. Discrete values are collected while the user is performing a tap gesture on the screen, during the action of taking pictures, used in the authentication or registration process. Both phases of the enhanced protocol employ the described deep neural networks and a binary classifier, respectively a Support Vector Machine (SVM).

We considered three options to enhance the ABC protocol with motion sensor fingerprints: using SVM classifiers on top of CNN embeddings, using SVM classifiers on top of ConvLSTM embeddings or using SVM classifiers on top of joint CNN and ConvLSTM embeddings. Throughout our experiments, we utilized two kernels for our SVM models, either linear or RBF, and two alternative values for the regularization parameter $C$, $10$ or $100$. For the RBF kernel, the parameter $\gamma$ is automatically scaled with respect to the number of features. The corresponding results are shown in Table~\ref{table_enhanced_abc}.

% , as shown in Table~\ref{table_enhanced_abc}. Hence, based on the generator used for the feature vectors, CNN, LSTM or a combination of the embeddings, the accuracy of our proposed extension varies. Further, our experiment results are outlined based on each embedding generator, and finally, we describe the results obtained from the concatenation of both methods. 

%We conducted experiments using the same set as in the previous experiments, using 6 smartphone devices and generating a total of 3000 attacks in order to provide a fair comparison between the proposed attack method and the enhanced ABC protocol. 

% - results from both abc + motion sensor for each type of model that produced feature embeddings
\subsubsection{Results with CNN Embeddings}

By extending the ABC protocol with motion sensor fingerprints based on CNN neural embeddings, we attain a top average accuracy of $99.47\%$ for the linear kernel and the regularization parameter $C\!=\!10$. In this case, the multi-modal ABC protocol successfully rejects our attack scheme with a false acceptance rate of $0.43\%$ and a false rejection rate of $1\%$, as shown in Table~\ref{table_enhanced_abc}. However, the smallest false acceptance ratio that can be obtained with CNN embeddings is $0.27\%$, using the RBF kernel and the regularization parameter $C=10$. In this case, the model is not very well balanced, the false rejection rate being $4.17\%$. We note that the accuracy of the multi-modal ABC protocol based on CNN embeddings is around $99\%$, irrespective of the kernel type or the regularization parameter value. Overall, the best SVM configuration for the CNN embeddings seems to be the one based on the linear kernel and the regularization $C=10$.

% \usepackage{graphicx}
\begin{table}[!t]
\renewcommand{\arraystretch}{1.2} 
\caption{Accuracy, FAR and FRR of the multi-modal ABC protocol for the attack scheme proposed in~\cite{benegui2020breach}, when five images and motion signals are used during registration and authentication. Results are reported for binary SVM classifiers based on neural embeddings from a CNN, a ConvLSTM or both. For each neural embedding type, we consider two kernels and different values for the regularization parameter $C$. All reported metrics represent average values computed for six smartphone devices, with $100$ authentic sessions and $500$ attack sessions per device. For each neural embedding type, the best results are highlighted in bold.}
\label{table_enhanced_abc}
\centering
\small{
\begin{tabular}{|l|c|c|c|c|}
\hline
\multicolumn{5}{|c|}{\textbf{ABC protocol + CNN embeddings}}  \\ \hline
\textbf{Kernel}   & \textbf{C}    & \textbf{Accuracy}  & \textbf{FAR} & \textbf{FRR}     \\ \hline
RBF               & 100           & 98.92\%            & 0.67\%         & 3.17\%  \\ \hline
RBF               & 10            & 99.08\%            & \textbf{0.27\%}         & 4.17\%  \\ \hline
Linear            & 100           & 98.97\%            & 0.80\%         & 2.17\%  \\ \hline
Linear            & 10            & \textbf{99.47\%}            & 0.43\%         & \textbf{1.00\%}  \\ \hline
\multicolumn{5}{|c|}{\textbf{ABC protocol + LSTM embeddings}} \\ \hline
\textbf{Kernel}   & \textbf{C}    & \textbf{Accuracy}  & \textbf{FAR} & \textbf{FRR}     \\ \hline
RBF               & 100           & 97.72\%            & 0.60\%         & 10.67\%  \\ \hline
RBF               & 10            & \textbf{99.03\%}            & \textbf{0.50\%}         & 3.33\%  \\ \hline
Linear            & 100           & 98.94\%            & 0.73\%         & \textbf{2.67\%}  \\ \hline
Linear            & 10            & \textbf{99.03\%}            & 0.53\%         & 3.17\%  \\ \hline
\multicolumn{5}{|c|}{\textbf{ABC protocol + CNN and LSTM embeddings}} \\ \hline
\textbf{Kernel}   & \textbf{C}    & \textbf{Accuracy}  & \textbf{FAR} & \textbf{FRR}     \\ \hline
RBF               & 100           & \textbf{99.67\%}            & \textbf{0.07\%}         & 1.67\%  \\ \hline
RBF               & 10            & 99.64\%            & 0.13\%         & \textbf{1.50\%}  \\ \hline
Linear            & 100           & 97.72\%            & 2.07\%         & 3.33\%  \\ \hline
Linear            & 10            & 99.39\%            & 0.13\%         & 3.00\%  \\ \hline
\end{tabular}%
}
\end{table}

\subsubsection{Results with ConvLSTM Embeddings}

The best accuracy attained by the multi-modal ABC protocol based on ConvLSTM embeddings is $99.03\%$, while the best false acceptance rate and the best false rejection rate are $0.50\%$ and $2.67\%$, respectively. However, these values are not attained with the same kernel and regularization parameter configuration. We are undecided regarding the best SVM configuration for the ConvLSTM embeddings. While the optimal regularization parameter is $C=10$, it appears that the linear and the RBF kernel produce equally good results for $C=10$. Another important remark is that the CNN embeddings seem to produce better motion sensor fingerprints than the ConvLSTM embeddings. With one exception (the configuration given by the RBF kernel and the regularization $C=100$), the differences in favor of the CNN embeddings are rather small.

\subsubsection{Results with Joint Neural Embeddings}

The results presented in Table~\ref{table_enhanced_abc} indicate that concatenating the neural embeddings generated by both CNN and ConvLSTM models gives superior performance levels than using the embeddings individually. By employing the RBF kernel and the regularization parameter $C=100$, we attain our lowest false acceptance rate of $0.07\%$ and our highest accuracy of $99.67\%$ in attack prevention. However, the lowest false rejection rate of $1.50\%$ is attained with the configuration based on the RBF kernel and the regularization parameter $C=10$. Overall, the joint CNN and ConvLSTM embeddings attain optimal results with the RBF kernel, irrespective of the value assigned to the regularization parameter.

\section{Conclusion}
\label{sec_Conclusion}

In this paper, we first presented an attack scheme for the ABC protocol proposed by Zhongjie et al.~\cite{Zhongjie-NDSS-2018}. Our attack scheme exposes a vulnerability in the original formulation of the ABC protocol, raising the false acceptance rate to $54.1\%$. Our attack scheme, which was initially proposed in~\cite{benegui2020breach}, is based on computing the attacker's camera fingerprint using photos taken outside the ABC protocol, allowing us to remove the fingerprint during impersonation attacks, without drastically altering the probe signal generated by the protocol. This procedure bypasses both protection systems of the original ABC protocol, namely Removal Detection and Forgery Detection.

Furthermore, we proposed a multi-modal ABC protocol based on deep neural networks applied on motion sensor signals, aiming to improve the security of the original ABC protocol. 
%Our enhanced ABC protocol utilizes discrete signals generated by motion sensors to identify the users using a binary classifier, namely an SVM. During the registration phase, motion sensor data collected during the tap gesture on the screen while the user is taking a photo, is used to assign a positive label ($+1$) to the user. Further, during the authentication phase in which the user takes photos of the two QR codes presented on the screen, we collect the motion data associated and send them to our SVM for binary classification. 
We processed the discrete signals using deep neural networks employed as feature extractors and we experimented with different kernels and embeddings in the meta-learning stage based on SVM. During the experiments, we identified that 512-dimensional neural embeddings, resulted from the concatenation of CNN and ConvLSTM embeddings, provided superior performance levels in attack prevention. Indeed, our multi-modal ABC protocol lowers the false acceptance rate for the attack proposed in~\cite{benegui2020breach} to as little as $0.07\%$, achieving, in the same time, a false rejection rate of $1.67\%$. Since motion sensor signals are not typically shared on social media platforms, we consider our multi-modal ABC protocol as much safer than the original formulation. Moreover, the users perform the exact same authentication steps in both original and multi-modal protocols. Hence, upgrading to the multi-modal ABC protocol does not imply additional authentication steps from the users.

One direction for future work is to turn our attention to transformer models \cite{Vaswani-NIPS-2017}. Transformers recently caught the attention of computer vision scientists \cite{dosovitskiy2021image}, as such models seem to have a better capacity of modeling global relations in the input. We believe that this property can lead to similar performance gains in motion signal processing. Another direction for future research is to adjust the proposed protocol to take advantage of the multiple cameras available on the high-end smartphone devices, which could provide a way to further enhance the protocol.

% or


% \appendices

% % you can choose not to have a title for an appendix
% % if you want by leaving the argument blank
% \section{}

%%%%%%%%%%%%%%%%%%%%%%%%%%%%%%%%%%%%%%%%%%
\vspace{6pt} 

%%%%%%%%%%%%%%%%%%%%%%%%%%%%%%%%%%%%%%%%%%
%% optional
%\supplementary{The following are available online at \linksupplementary{s1}, Figure S1: title, Table S1: title, Video S1: title.}


% %%%%%%%%%%%%%%%%%%%%%%%%%%%%%%%%%%%%%%%%%%
% \authorcontributions{For research articles with several authors, a short paragraph specifying their individual contributions must be provided. The following statements should be used ``Conceptualization, X.X. and Y.Y.; methodology, X.X.; software, X.X.; validation, X.X., Y.Y. and Z.Z.; formal analysis, X.X.; investigation, X.X.; resources, X.X.; data curation, X.X.; writing---original draft preparation, X.X.; writing---review and editing, X.X.; visualization, X.X.; supervision, X.X.; project administration, X.X.; funding acquisition, Y.Y. All authors have read and agreed to the published version of the manuscript.'', please turn to the  \href{http://img.mdpi.org/data/contributor-role-instruction.pdf}{CRediT taxonomy} for the term explanation. Authorship must be limited to those who have contributed substantially to the work~reported.}

\acknowledgments{The research leading to these results has received funding from the NO Grants 2014-2021, under project contract no. 24/2020. The article has also benefited from the support of the Romanian Young Academy, which is funded by Stiftung Mercator and the Alexander von Humboldt Foundation for the period 2020-2022.}

%%%%%%%%%%%%%%%%%%%%%%%%%%%%%%%%%%%%%%%%%%

\reftitle{References}


\externalbibliography{yes}
\bibliography{references}

\begin{thebibliography}{999}

\bibitem[{Board of Governors of the Federal Reserve
  System}(2016)]{Mobile-FRS-2016}
{Board of Governors of the Federal Reserve System}.
\newblock {Consumers and Mobile Financial Services 2016}.
\newblock
  \url{https://www.federalreserve.gov/econresdata/consumers-and-mobile-financial-services-report-201603.pdf},
   2016.
\newblock Accessed on: 2021-07-21.

\bibitem[Arthur(2013)]{arthur2013iphone}
Arthur, C.
\newblock {iPhone 5S fingerprint sensor hacked by Germany's Chaos Computer
  Club}.
\newblock {\em Guardian News. Np} {\bf 2013}, {\em 23}.

\bibitem[Aviv \em{et~al.}(2010)Aviv, Gibson, Mossop, Blaze, and
  Smith]{Aviv-WOOT-2010}
Aviv, A.J.; Gibson, K.; Mossop, E.; Blaze, M.; Smith, J.M.
\newblock {Smudge Attacks on Smartphone Touch Screens}.
\newblock  Proceedings of WOOT,  2010, pp. 1--7.

\bibitem[Xu \em{et~al.}(2013)Xu, Heinly, White, Monrose, and
  Frahm]{Xu-CCS-2013}
Xu, Y.; Heinly, J.; White, A.M.; Monrose, F.; Frahm, J.M.
\newblock {Seeing double: Reconstructing obscured typed input from repeated
  compromising reflections}.
\newblock  Proceedings of CCS,  2013, pp. 1063--1074.

\bibitem[Zhang \em{et~al.}(2012)Zhang, Xia, Luo, Ling, Liu, and
  Fu]{Zhang-SPSM-2012}
Zhang, Y.; Xia, P.; Luo, J.; Ling, Z.; Liu, B.; Fu, X.
\newblock Fingerprint attack against touch-enabled devices.
\newblock  Proceedings of SPSM,  2012, pp. 57--68.

\bibitem[Zhongjie \em{et~al.}(2018)Zhongjie, Sixu, Xinwen, Dimitrios, Aziz, and
  Kui]{Zhongjie-NDSS-2018}
Zhongjie, B.; Sixu, P.; Xinwen, F.; Dimitrios, K.; Aziz, M.; Kui, R.
\newblock {ABC: Enabling Smartphone Authentication with Built-in Camera}.
\newblock  Proceedings of NDSS,  2018.

\bibitem[Luk{\'a}{\v{s}} \em{et~al.}(2006)Luk{\'a}{\v{s}}, Fridrich, and
  Goljan]{Lukas-TIFS-2006}
Luk{\'a}{\v{s}}, J.; Fridrich, J.; Goljan, M.
\newblock {Digital camera identification from sensor pattern noise}.
\newblock {\em IEEE Transactions on Information Forensics and Security} {\bf
  2006}, {\em 1},~205--214.

\bibitem[Altinisik \em{et~al.}(2018)Altinisik, Tasdemir, and
  Sencar]{Altinisik-EUSIPCO-2018}
Altinisik, E.; Tasdemir, K.; Sencar, H.T.
\newblock {Extracting PRNU Noise from H.264 Coded Videos}.
\newblock  Proceedings of EUSIPCO,  2018, pp. 1367--1371.

\bibitem[Akshatha \em{et~al.}(2016)Akshatha, Karunakar, Anitha, Raghavendra,
  and Shetty]{Akshatha-DI-2016}
Akshatha, K.; Karunakar, A.; Anitha, H.; Raghavendra, U.; Shetty, D.
\newblock {Digital camera identification using PRNU: A feature based approach}.
\newblock {\em Digital Investigation} {\bf 2016}, {\em 19},~69--77.

\bibitem[Li(2010)]{Li-TIFS-2010}
Li, C.T.
\newblock Source camera identification using enhanced sensor pattern noise.
\newblock {\em IEEE Transactions on Information Forensics and Security} {\bf
  2010}, {\em 5},~280--287.

\bibitem[Cooper(2013)]{Cooper-FSI-2013}
Cooper, A.J.
\newblock {Improved photo response non-uniformity (PRNU) based source camera
  identification}.
\newblock {\em Forensic Science International} {\bf 2013}, {\em 226},~132--141.

\bibitem[Kang \em{et~al.}(2012)Kang, Li, Qu, and Huang]{Kang-TIFS-2012}
Kang, X.; Li, Y.; Qu, Z.; Huang, J.
\newblock Enhancing source camera identification performance with a camera
  reference phase sensor pattern noise.
\newblock {\em IEEE Transactions on Information Forensics and Security} {\bf
  2012}, {\em 7},~393--402.

\bibitem[Benegui and Ionescu(2020)]{benegui2020breach}
Benegui, C.; Ionescu, R.T.
\newblock {A breach into the Authentication with Built-in Camera (ABC)
  Protocol}.
\newblock  Proceedings of ACNS,  2020, pp. 3--20.

\bibitem[LeCun \em{et~al.}(1998)LeCun, Bottou, Bengio, and
  Haffner]{LeCun-IEEE-1998}
LeCun, Y.; Bottou, L.; Bengio, Y.; Haffner, P.
\newblock {Gradient-based Learning Applied to Document Recognition}.
\newblock {\em Proceedings of the IEEE} {\bf 1998}, {\em 86},~2278--2324.

\bibitem[Krizhevsky \em{et~al.}(2012)Krizhevsky, Sutskever, and
  Hinton]{Krizhevsky-NIPS-2012}
Krizhevsky, A.; Sutskever, I.; Hinton, G.E.
\newblock {ImageNet Classification with Deep Convolutional Neural Networks}.
\newblock  Proceedings of NIPS,  2012, pp. 1097--1105.

\bibitem[Xingjian \em{et~al.}(2015)Xingjian, Chen, Wang, Yeung, Wong, and
  Woo]{Xingjian-NIPS-2015}
Xingjian, S.; Chen, Z.; Wang, H.; Yeung, D.Y.; Wong, W.K.; Woo, W.C.
\newblock {Convolutional LSTM network: A machine learning approach for
  precipitation nowcasting}.
\newblock  Proceedings of NIPS,  2015, pp. 802--810.

\bibitem[Cortes and Vapnik(1995)]{Cortes-ML-1995}
Cortes, C.; Vapnik, V.
\newblock {Support-Vector Networks}.
\newblock {\em Machine Learning} {\bf 1995}, {\em 20},~273--297.

\bibitem[Aghili and Mala(2018)]{Aghili-IACR-2018}
Aghili, S.F.; Mala, H.
\newblock {Breaking a Lightweight {M2M} Authentication Protocol for
  Communications in IIoT Environment}.
\newblock {\em IACR Cryptology ePrint Archive} {\bf 2018}, {\em 2018},~891.

\bibitem[Esfahani \em{et~al.}(2019)Esfahani, Mantas, Matischek, Saghezchi,
  Rodriguez, Bicaku, Maksuti, Tauber, Schmittner, and
  Bastos]{Esfahani-ITJ-2019}
Esfahani, A.; Mantas, G.; Matischek, R.; Saghezchi, F.B.; Rodriguez, J.;
  Bicaku, A.; Maksuti, S.; Tauber, M.; Schmittner, C.; Bastos, J.
\newblock {A lightweight authentication mechanism for M2M communications in
  industrial IoT environment}.
\newblock {\em IEEE Internet of Things Journal} {\bf 2019}, {\em 6},~288--296.

\bibitem[Aghili \em{et~al.}(2018)Aghili, Mala, and
  Peris{-}Lopez]{Aghili-Sensors-2018}
Aghili, S.F.; Mala, H.; Peris{-}Lopez, P.
\newblock {Securing Heterogeneous Wireless Sensor Networks: Breaking and Fixing
  a Three-Factor Authentication Protocol}.
\newblock {\em Sensors} {\bf 2018}, {\em 18},~3663.

\bibitem[Amin \em{et~al.}(2018)Amin, Islam, Kumar, and Choo]{Amin-JNCA-2018}
Amin, R.; Islam, S.H.; Kumar, N.; Choo, K.K.R.
\newblock An untraceable and anonymous password authentication protocol for
  heterogeneous wireless sensor networks.
\newblock {\em Journal of Network and Computer Applications} {\bf 2018}, {\em
  104},~133--144.

\bibitem[Amerini \em{et~al.}(2016)Amerini, Bestagini, Bondi, Caldelli, Casini,
  and Tubaro]{DBLP:conf/mediaforensics/AmeriniBBCCT16}
Amerini, I.; Bestagini, P.; Bondi, L.; Caldelli, R.; Casini, M.; Tubaro, S.
\newblock Robust smartphone fingerprint by mixing device sensors features for
  mobile strong authentication.
\newblock  Media Watermarking, Security, and Forensics. Ingenta,  2016, pp.
  1--8.

\bibitem[Valsesia \em{et~al.}(2017)Valsesia, Coluccia, Bianchi, and
  Magli]{DBLP:journals/tifs/ValsesiaCBM17}
Valsesia, D.; Coluccia, G.; Bianchi, T.; Magli, E.
\newblock {User Authentication via PRNU-Based Physical Unclonable Functions}.
\newblock {\em IEEE Transactions on Information Forensics and Security} {\bf
  2017}, {\em 12},~1941--1956.

\bibitem[Gloe \em{et~al.}(2007)Gloe, Kirchner, Winkler, and
  B{\"o}hme]{gloe2007can}
Gloe, T.; Kirchner, M.; Winkler, A.; B{\"o}hme, R.
\newblock Can we trust digital image forensics?
\newblock  Proceedings of ACMMM,  2007, pp. 78--86.

\bibitem[Goljan \em{et~al.}(2011)Goljan, Fridrich, and
  Chen]{goljan2011defending}
Goljan, M.; Fridrich, J.; Chen, M.
\newblock Defending against fingerprint-copy attack in sensor-based camera
  identification.
\newblock {\em IEEE Transactions on Information Forensics and Security} {\bf
  2011}, {\em 6},~227--236.

\bibitem[Buriro \em{et~al.}(2016)Buriro, Crispo, Delfrari, and
  Wrona]{Buriro-SPW-2016}
Buriro, A.; Crispo, B.; Delfrari, F.; Wrona, K.
\newblock Hold and Sign: A Novel Behavioral Biometrics for Smartphone User
  Authentication.
\newblock  Proceedings of SPW,  2016, pp. 276--285.

\bibitem[Buriro \em{et~al.}(2017)Buriro, Crispo, and
  Zhauniarovich]{Buriro-ISBA-2017}
Buriro, A.; Crispo, B.; Zhauniarovich, Y.
\newblock {Please Hold On: Unobtrusive User Authentication using Smartphone's
  built-in Sensors}.
\newblock  Proceedings of ISBA,  2017, pp. 1--8.

\bibitem[Ku \em{et~al.}(2019)Ku, Park, Shin, and Kwon]{Ku-Access-2019}
Ku, Y.; Park, L.H.; Shin, S.; Kwon, T.
\newblock Draw It As Shown: Behavioral Pattern Lock for Mobile User
  Authentication.
\newblock {\em IEEE Access} {\bf 2019}, {\em 7},~69363--69378.

\bibitem[Li \em{et~al.}(2018)Li, Yu, and Cao]{Li-BIBM-2018}
Li, H.; Yu, J.; Cao, Q.
\newblock {Intelligent Walk Authentication: Implicit Authentication When You
  Walk with Smartphone}.
\newblock  Proceedings of BIBM,  2018, pp. 1113--1116.

\bibitem[Neverova \em{et~al.}(2016)Neverova, Wolf, Lacey, Fridman, Chandra,
  Barbello, and Taylor]{Neverova-Access-2016}
Neverova, N.; Wolf, C.; Lacey, G.; Fridman, L.; Chandra, D.; Barbello, B.;
  Taylor, G.
\newblock {Learning Human Identity from Motion Patterns}.
\newblock {\em IEEE Access} {\bf 2016}, {\em 4},~1810--1820.

\bibitem[Shen \em{et~al.}(2016)Shen, Yu, Yuan, Li, and Guan]{Shen-Sensors-2016}
Shen, C.; Yu, T.; Yuan, S.; Li, Y.; Guan, X.
\newblock {Performance Analysis of Motion-Sensor Behavior for User
  Authentication on Smartphones}.
\newblock {\em Sensors} {\bf 2016}, {\em 16},~345.

\bibitem[Sitov{\'a} \em{et~al.}(2016)Sitov{\'a}, {\v{S}}edenka, Yang, Peng,
  Zhou, Gasti, and Balagani]{Sitova-TIFS-2016}
Sitov{\'a}, Z.; {\v{S}}edenka, J.; Yang, Q.; Peng, G.; Zhou, G.; Gasti, P.;
  Balagani, K.S.
\newblock {HMOG: New Behavioral Biometric Features for Continuous
  Authentication of Smartphone Users}.
\newblock {\em IEEE Transactions on Information Forensics and Security} {\bf
  2016}, {\em 11},~877--892.

\bibitem[Sun \em{et~al.}(2017)Sun, Wang, Cao, Philip, Srisa-An, and
  Leow]{Sun-ECML-2017}
Sun, L.; Wang, Y.; Cao, B.; Philip, S.Y.; Srisa-An, W.; Leow, A.D.
\newblock Sequential keystroke behavioral biometrics for mobile user
  identification via multi-view deep learning.
\newblock  Proceedings of ECML-PKDD,  2017, pp. 228--240.

\bibitem[Vildjiounaite \em{et~al.}(2006)Vildjiounaite, M{\"a}kel{\"a},
  Lindholm, Riihim{\"a}ki, Kyll{\"o}nen, M{\"a}ntyj{\"a}rvi, and
  Ailisto]{Vildjiounaite-ICPC-2006}
Vildjiounaite, E.; M{\"a}kel{\"a}, S.M.; Lindholm, M.; Riihim{\"a}ki, R.;
  Kyll{\"o}nen, V.; M{\"a}ntyj{\"a}rvi, J.; Ailisto, H.
\newblock Unobtrusive multimodal biometrics for ensuring privacy and
  information security with personal devices.
\newblock  Proceedings of PERVASIVE,  2006, pp. 187--201.

\bibitem[Wang and Tao(2019)]{Wang-Access-2019}
Wang, R.; Tao, D.
\newblock {Context-Aware Implicit Authentication of Smartphone Users Based on
  Multi-Sensor Behavior}.
\newblock {\em IEEE Access} {\bf 2019}, {\em 7},~119654--119667.

\bibitem[Benegui and Ionescu(2020)]{Benegui-Access-2020}
Benegui, C.; Ionescu, R.T.
\newblock {Convolutional Neural Networks for User Identification based on
  Motion Sensors Represented as Images}.
\newblock {\em IEEE Access} {\bf 2020}, {\em 8},~61255--61266.

\bibitem[Goljan(2008)]{Goljan-IWDW-2008}
Goljan, M.
\newblock Digital camera identification from images -- estimating false
  acceptance probability.
\newblock  Proceedings of IWDW,  2008, pp. 454--468.

\bibitem[Simonyan and Zisserman(2014)]{Simonyan-ICLR-14}
Simonyan, K.; Zisserman, A.
\newblock {Very Deep Convolutional Networks for Large-Scale Image Recognition}.
\newblock  Proceedings of ICLR,  2014.

\bibitem[He \em{et~al.}(2016)He, Zhang, Ren, and Sun]{He-CVPR-2016}
He, K.; Zhang, X.; Ren, S.; Sun, J.
\newblock {Deep Residual Learning for Image Recognition}.
\newblock  Proceedings of CVPR,  2016, pp. 770--778.

\bibitem[Srivastava \em{et~al.}(2014)Srivastava, Hinton, Krizhevsky, Sutskever,
  and Salakhutdinov]{Srivastava-JMLR-2014}
Srivastava, N.; Hinton, G.; Krizhevsky, A.; Sutskever, I.; Salakhutdinov, R.
\newblock {Dropout: A Simple Way to Prevent Neural Networks from Overfitting}.
\newblock {\em Journal of Machine Learning Research} {\bf 2014}, {\em
  15},~1929--1958.

\bibitem[Nair and Hinton(2010)]{Nair-ICML-2010}
Nair, V.; Hinton, G.E.
\newblock {Rectified Linear Units Improve Restricted Boltzmann Machines}.
\newblock  Proceedings of ICML,  2010, pp. 807--814.

\bibitem[Kingma and Ba(2015)]{Kingma-ICLR-2015}
Kingma, D.P.; Ba, J.
\newblock Adam: A method for stochastic optimization.
\newblock  Proceedings of ICLR,  2015.

\bibitem[Quiring and Kirchner(2015)]{Quiring-WIFS-2015}
Quiring, E.; Kirchner, M.
\newblock {Fragile sensor fingerprint camera identification}.
\newblock  Proceedings of WIFS,  2015, pp. 1--6.

\bibitem[Vaswani \em{et~al.}(2017)Vaswani, Shazeer, Parmar, Uszkoreit, Jones,
  Gomez, Kaiser, and Polosukhin]{Vaswani-NIPS-2017}
Vaswani, A.; Shazeer, N.; Parmar, N.; Uszkoreit, J.; Jones, L.; Gomez, A.N.;
  Kaiser, {\L}.; Polosukhin, I.
\newblock {Attention is All You Need}.
\newblock  Proceedings of NIPS,  2017, pp. 5998--6008.

\bibitem[Dosovitskiy \em{et~al.}(2021)Dosovitskiy, Beyer, Kolesnikov,
  Weissenborn, Zhai, Unterthiner, Dehghani, Minderer, Heigold, Gelly,
  Uszkoreit, and Houlsby]{dosovitskiy2021image}
Dosovitskiy, A.; Beyer, L.; Kolesnikov, A.; Weissenborn, D.; Zhai, X.;
  Unterthiner, T.; Dehghani, M.; Minderer, M.; Heigold, G.; Gelly, S.;
  Uszkoreit, J.; Houlsby, N.
\newblock {An Image is Worth 16x16 Words: Transformers for Image Recognition at
  Scale}.
\newblock  Proceedings of ICLR,  2021.

\end{thebibliography}

% If authors have biography, please use the format below
%\section*{Short Biography of Authors}
%\bio
%{\raisebox{-0.35cm}{\includegraphics[width=3.5cm,height=5.3cm,clip,keepaspectratio]{Definitions/author1.pdf}}}
%{\textbf{Firstname Lastname} Biography of first author}
%
%\bio
%{\raisebox{-0.35cm}{\includegraphics[width=3.5cm,height=5.3cm,clip,keepaspectratio]{Definitions/author2.jpg}}}
%{\textbf{Firstname Lastname} Biography of second author}

% The following MDPI journals use author-date citation: Arts, Econometrics, Economies, Genealogy, Humanities, IJFS, JRFM, Laws, Religions, Risks, Social Sciences. For those journals, please follow the formatting guidelines on http://www.mdpi.com/authors/references
% To cite two works by the same author: \citeauthor{ref-journal-1a} (\citeyear{ref-journal-1a}, \citeyear{ref-journal-1b}). This produces: Whittaker (1967, 1975)
% To cite two works by the same author with specific pages: \citeauthor{ref-journal-3a} (\citeyear{ref-journal-3a}, p. 328; \citeyear{ref-journal-3b}, p.475). This produces: Wong (1999, p. 328; 2000, p. 475)

%%%%%%%%%%%%%%%%%%%%%%%%%%%%%%%%%%%%%%%%%%
%% for journal Sci
%\reviewreports{\\
%Reviewer 1 comments and authors’ response\\
%Reviewer 2 comments and authors’ response\\
%Reviewer 3 comments and authors’ response
%}
%%%%%%%%%%%%%%%%%%%%%%%%%%%%%%%%%%%%%%%%%%
\end{paracol}
\end{document}